# Magnetic anisotropy of the single-crystalline ferromagnetic insulator Cr$_2$Ge$_2$Te$_6$

Xiao Zhang[1,2], Yuelei Zhao[1,2], Qi Song[1,2], Shuang Jia[1,2], Jing Shi[3*], and Wei Han[1,2*]

[1]International Center for Quantum Materials, School of Physics, Peking University, Beijing 100871, P. R. China

[2]Collaborative Innovation Center of Quantum Matter, Beijing 100871, P. R. China

[3]Department of Physics and Astronomy, University of California, Riverside, CA 92521, USA

*Email: jing.shi@ucr.edu; weihan@pku.edu.cn

**Abstract:**

Cr$_2$Ge$_2$Te$_6$ (CGT), a layered ferromagnetic insulator, has attracted a great deal of interest recently owing to its potential for integration with Dirac materials to realize the quantum anomalous Hall effect (QAHE) and to develop novel spintronics devices. Here, we study the uniaxial magnetic anisotropy energy of single-crystalline CGT and determine that the magnetic easy axis is directed along the c-axis in its ferromagnetic phase. In addition, CGT is an insulator below the Curie temperature. These properties make CGT a potentially promising candidate substrate for integration with topological insulators for the realization of the high-temperature QAHE.

## 1. Introduction

There has been enormous interest in realizing the quantum anomalous Hall effect (QAHE) by introducing a magnetic exchange gap into the Dirac spectrum of surface states of topological insulators (TI) [1-8]. One promising route is to directly exchange couple the TI surface states to a ferromagnetic insulator (FI) via the proximity effect, which does not resort to doping the TI with magnetic elements that could unavoidably increase the disorder [9-12]. $Cr_2Ge_2Te_6$ (CGT), one of a few known ferromagnetic insulators, is particularly interesting because of its relatively high Curie temperature (~ 61 K) [13]. In addition, the epitaxial growth of $Bi_2Te_3$ on CGT [14] and magnetic order in the heterostructures of $Bi_2Se_3$/CGT has been reported recently [10]. In those reports [10,13,14], the electrical properties, the magnetization, and Curie temperature have been intensively studied. For the applications of the QAHE, the magnetic anisotropy perpendicular to the TI films is favored. Thus, it is necessary to study the magnetic anisotropic properties of CGT.

In this letter, we report the characterization of the uniaxial magnetic anisotropy energy of the single-crystalline layered CGT synthesized by the flux method. The electrical properties are first characterized; CGT is insulating at low temperatures (T <100 K). The Curie temperature is determined to be ~ 61 K. Furthermore, we determine that the magnetic easy axis of CGT is directed along the c-axis in its ferromagnetic phase, from both ferromagnetic resonance and magnetization measurements.

## 2. Experimental procedure

Single-crystalline layered CGT is synthesized by the flux method from high-purity elemental Cr (99.99%), Ge (99.999%), and Te (>99.999%) materials obtained from Alfa Aesar [13]. The procedure to prepare the single crystal is as follows. Firstly, the initial atomic ratio of Cr:Ge:Te is 10:13.5:76.5, where the extra Ge and Te are used as fluxes. Then, these materials are sealed in evacuated quartz ampoules and kept at 1000 °C for 1 h, followed by slow cooling to 450 °C over a long period of 90 h. After removing the extra Ge and Te by a centrifugation step, the extra Te is evaporated in a sealed evacuated quartz ampoule at 450 °C for 48 hours in a horizontal tube furnace with a lateral temperature gradient, following the procedure described in an earlier report [14].

The crystalline structural properties of CGT samples are characterized using a powder X-ray diffractometer and then the results are refined using a Rietica Rietveld program. The electrical transport properties of CGT are characterized by measuring the temperature dependence of the resistivity from 300 to 100 K in a closed-cycle refrigerator. The magnetic properties of CGT are measured using the Magnetic Properties Measurement System (MPMS-3; Quantum Design), and the ferromagnetic resonance of CGT is measured using a vector network analyzer (VNA; Agilent E5071C) connected with a coplanar waveguide in the variable-temperature insert of a Quantum Design Physical Properties Measurement System (PPMS; Quantum Design) .

3. Result and discussion

The X-ray diffraction (XRD) pattern of CGT is shown in Fig. 1(a); the observed peaks fit well in the $R\bar{3}$ group. Moreover, the Ge and Te impurities are undetectable, indicating that approximately all the Ge and Te fluxes are removed. The synthesized CGT single crystals are typically 3 to 4 mm in size, as shown in Fig. 1(b). Fig. 1(c) shows the schematic drawing of the crystal structure of CGT. Furthermore, we measure the XRD from the CGT crystal's ab plane. As shown in Fig. 1(d), only (003), (006), and (0012) peaks are present, indicating the single-crystalline quality of the CGT samples in our studies.

The electrical transport properties of CGT are characterized by measuring the temperature dependence of the resistivity from 300 to 100 K. The sample is 1.5 mm wide and 3.0 mm long with four silver paint contacts equally spaced on the surface. During the measurement, the current is kept at 0.1 µA and the voltage is measured with a nanovoltage meter. The temperature-dependent resistivity indicates a semiconductor-like behavior, as shown in Fig. 2(a). At room temperature, the resistivity is ~ 0.42 Ω cm, and it monotonically increases as the temperature decreases. When the temperature decreases to 125 K, we observe a very sharp increase in the resistivity and it becomes ~ 2000 Ω cm at around 100 K. Below 100 K, we are not able to obtain the data because the sample resistivity is too high to measure. These observations indicate the good insulating property of our single-crystal CGT sample at low temperatures. Further reduction of the parasitic conduction in the CGT substrate is required to realize the QAHE. Two possible routes are to improve the purity of the CGT crystal to decrease unwanted carriers and to fabricate CGT substrates as thin as several atom layers by mechanical exfoliation. The temperature-dependent resistivity can be fitted to the thermal activation model with a constant energy gap, $E_a$, as described by

$$\rho = \rho_0 \cdot \exp(\frac{E_a}{2k_B T}) \ , \qquad (1)$$

where $k_B$ is the Boltzmann constant and $T$ is the temperature. To obtain the value of the transport energy gap, $E_a$, we plot the natural logarithm of the resistivity (in the unit of Ω cm) versus $1/T$, as shown in the inset of Fig. 2(a). From the linearly fitted curve (red line), $E_a$ is obtained to be ~0.24 eV, which is similar to a previously reported value of ~ 0.2 eV [14].

To identify the Curie temperature, $T_C$, we first measure the magnetization of the CGT sample from 5 to 300 K with a magnetic field of 0.1 T parallel to the crystal's ab plane [green line in Fig. 2(b)]. The magnetization shows an abrupt decrease at about ~ 61 K, as indicated by the red arrow. The obtained $T_C$ is ~ 61 K, which is in good agreement with the values reported previously [13,14]. Below $T_C$, the magnetization shows a modest decrease as the temperature decreases. This could be attributed to the fact that the magnetization does not reach its saturation under the magnetic field of 0.1 T owing to increased magnetic anisotropy, which is discussed below.

To confirm this Curie temperature, we also measured the magnetization of the CGT sample from 5 to 300 K with a magnetic field of 2 T parallel to the crystal's c axis, as shown in Fig. 2(b) (black line). It is well known that the temperature-dependent magnetic susceptibility follows the Curie-Weiss law:

$$\chi = \frac{M}{H} = \frac{C}{T - T_C}, \qquad (2)$$

where $\chi$ is the magnetic susceptibility, M is the magnetization, and H is the magnetic field strength. C is a material-specific Curie constant. To obtain the value of Curie temperature, we plot the reciprocal of $\chi$ versus $T$, as shown in the inset of Fig. 2(b). On the basis of the linear fitted curve (red line), $T_C$ is obtained to be ~ 70 K, which is slightly higher than our observable $T_C$ ~ 61K with an unsaturated magnetic field.

The magnetic anisotropy of CGT is first characterized by ferromagnetic resonance (FMR) [15]. The temperature range is from 5 to 300 K and the magnetic field is applied parallel to the crystal's *ab* plane. Figure 3(a) shows four representative curves of the forward amplitudes of the complex transmission coefficients ($S_{21}$) vs magnetic field measured at the frequencies of 5, 7, 9, and 11

GHz at 60 K after renormalization by subtracting a constant background. To determine the magnetic anisotropy, $S_{21}$ is measured as a function of both frequency and magnetic field for various temperatures. The results measured at 60, 30, and 5 K are shown in Figs. 3(b)-3(d), where the red and blue colors represent high and low transmissions. The FMR signal is clearly observable in the ferromagnetic phase, but not easily identifiable in the paramagnetic phase. This observation is consistent with the Curie temperature determined from the measurement of the temperature-dependent magnetization.

There is an obvious minimum feature for the resonance frequency, as indicated by the black arrows in Figs. 3(b)-3(d). This observation could be attributed to the magnetic anisotropy between the ab-plane and the c axis owing to CGT's layered structure. In our measurement, the resonance field where this minimum feature occurs, $H_{tran}$, could be related to the magnetic anisotropy in CGT samples by the following equation [16]:

$$H_{tran} = \frac{2K_u}{M_S}, \tag{3}$$

where $K_u$ is the uniaxial magnetic anisotropy energy along the c-axis and $M_s$ is the saturation magnetization. From the magnetization measured at various temperatures [Figs. 4(a)-4(c)], we calculated the temperature dependence of $K_u$ [Fig. 3(e)]. $K_u$ is found to be in the range of (1.35 - 3.65) × $10^5$ erg/cm$^3$ from 5 to 60 K, which indicates that the magnetic easy axis of CGT is perpendicular to the crystal's *ab* plane in its ferromagnetic phase.

To further confirm the magnetic anisotropy of CGT, the magnetization vs magnetic field is measured at various temperatures [Figs. 4(a)-4(c)]. During the measurement, the magnetic field is applied either parallel to the crystal's *ab* plane or the crystal's *c* axis. When the field is parallel to the crystal's *c* axis, the magnetic field needed for the full saturation of CGT's magnetization is much lower, indicating that the magnetic easy axis is along the crystal's *c* axis. At 5 K, the magnetization shows a dramatic increase at ~ 0.5 T and then reaches saturation at the value ~2.23 $\mu_B$ / Cr atom. This value is slightly lower than the expected value for the high spin configuration state of $Cr^{3+}$ (3 $\mu_B$/Cr atom), which might be caused by some nonmagnetic impurities in this sample.

When the magnetic field is applied parallel to the crystal's *ab* plane during the measurement, the saturation magnetic field, $H_{sat}$, is associated with the uniaxial magnetic anisotropic energy and the magnetization, as shown in the equation below [17]:

$$K_u = \frac{1}{2} H_{sat} \cdot M_S. \qquad (4)$$

The $K_u$ is calculated to be in the range of (1.27 - 3.66) × $10^5$ erg/cm$^3$ from 5 to 60 K, as summarized in Fig. 4(d). These values are putatively consistent with the values determined from the FMR measurements, and indicate that the magnetic easy axis of CGT is perpendicular to the crystal's *ab* plane in its ferromagnetic phase.

## 4. Conclusions

The single-crystalline CGT samples exhibit a magnetic easy axis along the crystal's *c* axis, as determined from both FMR and magnetization measurements. This perpendicular magnetic anisotropy and the insulating property below Curie temperature will make CGT one of the most promising candidate substrates for integration with TIs for the realization of the high-temperature QAHE.

**Acknowledgements**

We acknowledge the technical help and discussion from Jiarui Li, Wei Yuan, and Zeji Wang, and the funding support of the National Basic Research Programs of China (973 Grants 2013CB921903, 2015CB921104, 2014CB920902, 2013CB921901, and 2014CB239302) and the National Natural Science Foundation of China (NSFC Grant 11574006). Wei Han also acknowledges the support of the 1000 Talents Program for Young Scientists of China.

**Figure Captions**

Figure 1. (Color online) Structural characteristics of $Cr_2Ge_2Te_6$ (CGT). (a) Experimental (black crosses), and calculated (red line) diffractograms for a typical CGT sample. The green line indicates the difference between the experimental results and the calculation, and the short dashed blue lines indicate the expected peak positions for the $R\bar{3}$ group of the single crystalline CGT. (b) Image of a representative single-crystalline CGT sample synthesized by the flux method. (c) Schematic drawing of the one-unit cell for CGT single crystal. (d) X-ray diffraction pattern from the crystal's *ab* plane of CGT.

Figure 2. (Color online) Temperature-dependent electrical and magnetic transport properties of the single-crystalline CGT. (a) Resistivity versus temperature from 300 to 100 K. Inset: Linear fit (red line) of the logarithmic plot of resistivity as a function of the reciprocal of temperature. (b) Temperature dependence of magnetization with an applied field of 0.1 T parallel to the crystal's

*ab* plane and 2 T parallel to the crystal's *c* axis. Inset: Data (black blocks) and linear fit (red line) of the reciprocal plot of magnetic susceptibility as a function of temperature for 2 T, H//c axis.

Figure 3. (Color online) Ferromagnetic resonance measurement of CGT. (a) $S_{21}$, the forward amplitude of the complex transmission coefficient, versus magnetic field at 60 K. Different colors represent signals with different frequencies. (b-d) Forward amplitudes of the complex transmission coefficient as functions of magnetic field, $H$, and microwave frequency, $f$. The magnitude is indicated by the same color code for all three figures. The magnetic field is applied parallel to the crystal's *ab* plane. (e) The uniaxial magnetic anisotropic energy, $K_u$, calculated from ferromagnetic resonance, as a function of temperature.

Figure 4. (Color online) Field dependent magnetization of CGT measured with Magnetic Properties Measurement System. (a-c) Magnetization curves measured at 5, 30, and 60 K, respectively. The magnetic field is applied parallel to the crystal's *ab* plane. Insets: Magnetic-field-dependent magnetization when magnetic field is parallel to *c* axis. (d) Uniaxial magnetic anisotropic energy, $K_u$, as a function of temperature.

Figure 1

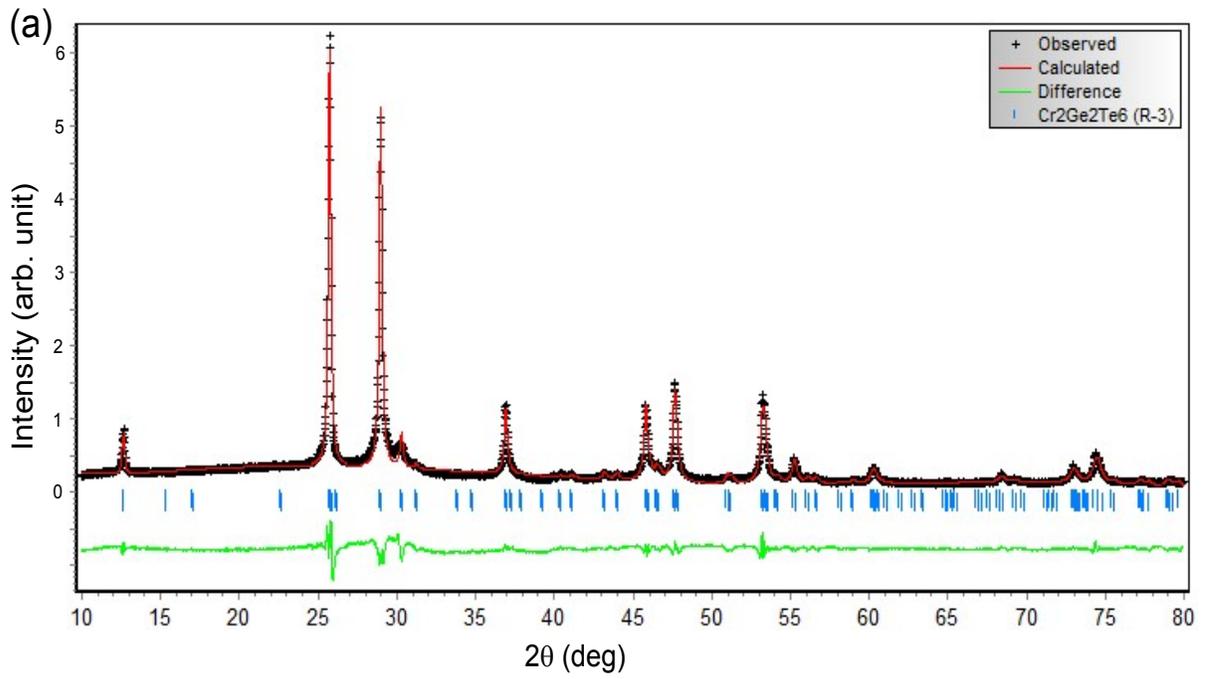
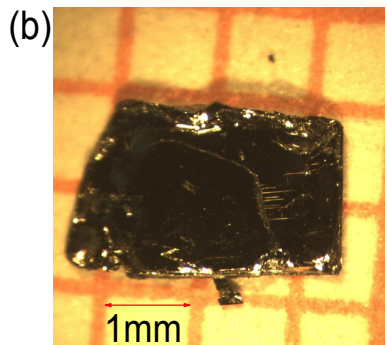
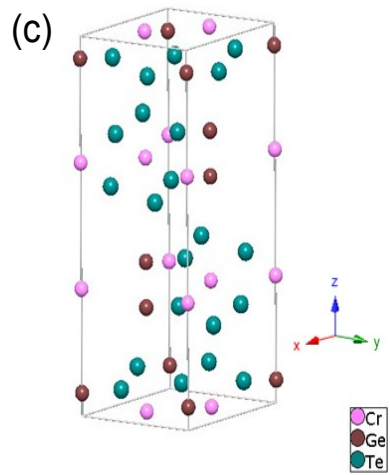
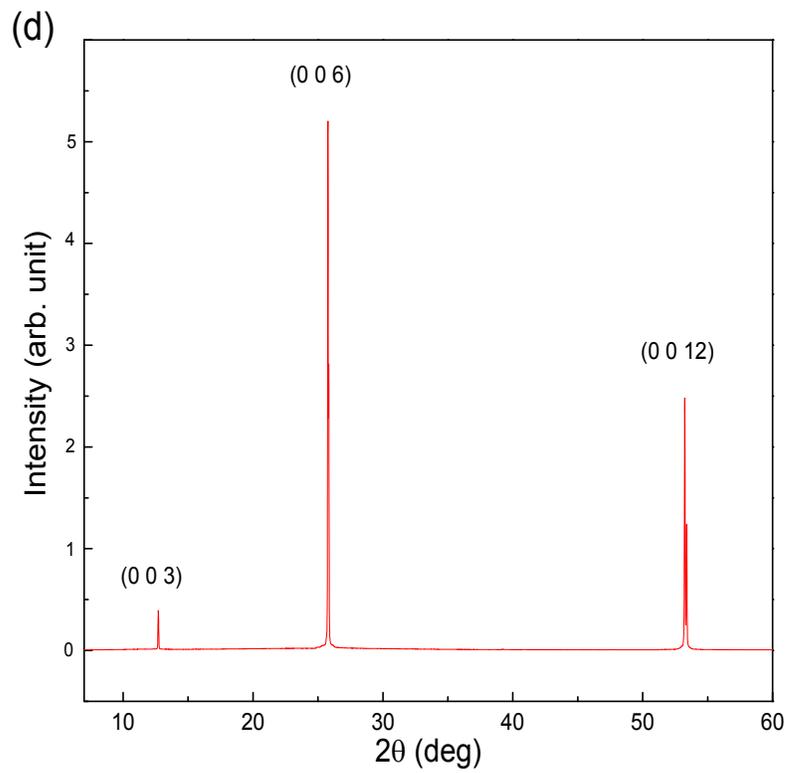

Figure 2

(a)
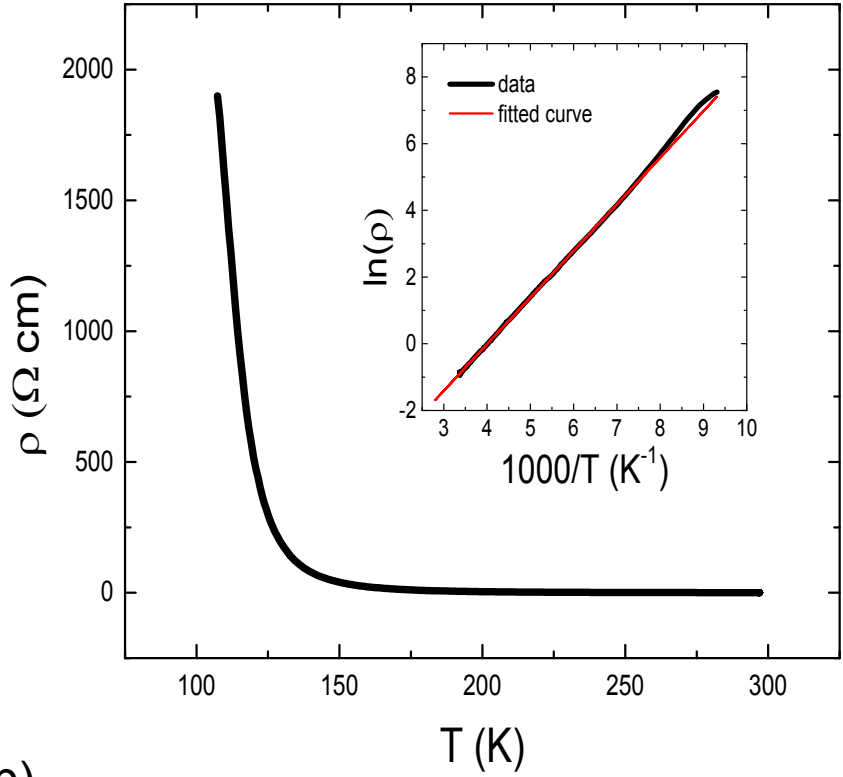

(b)
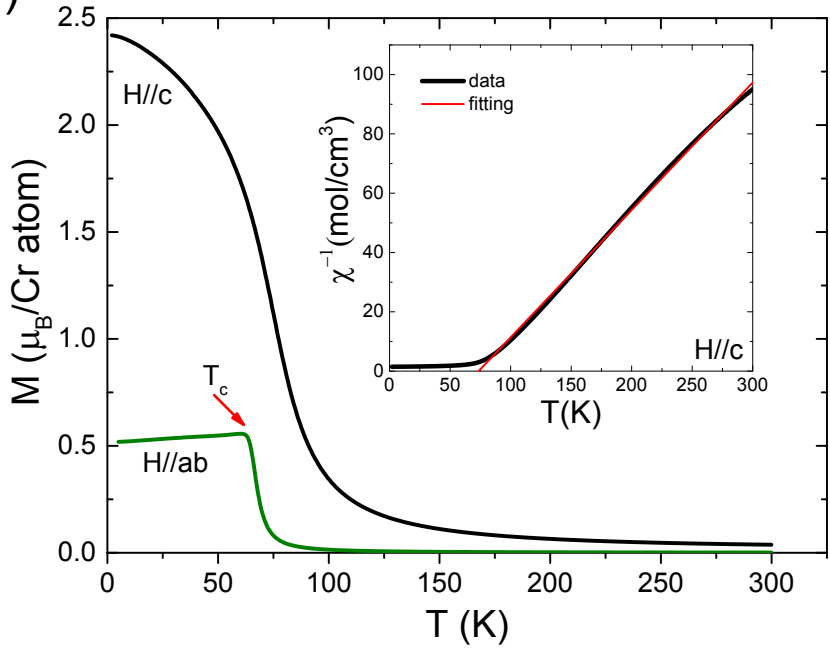

Figure 3

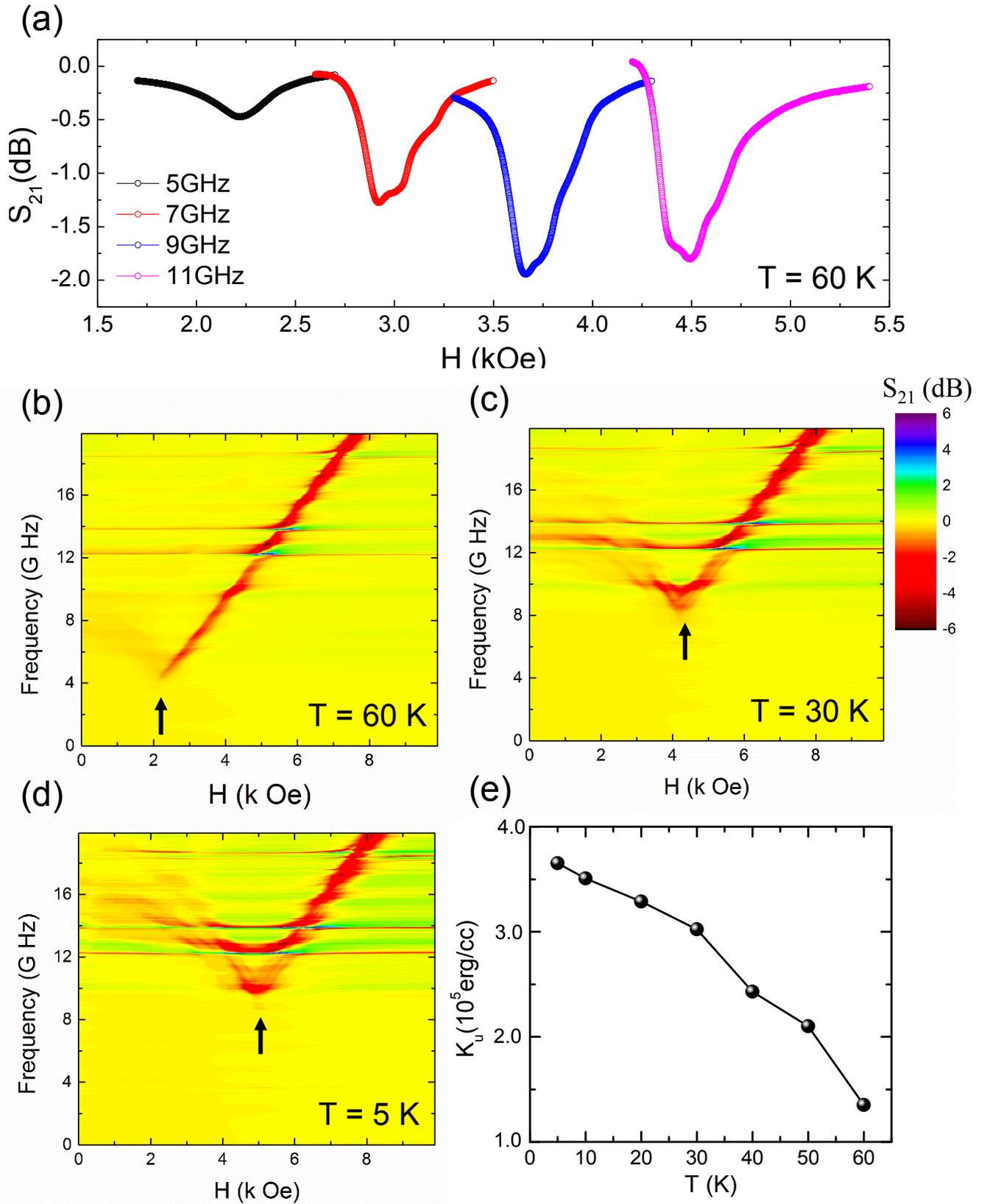

Figure 4

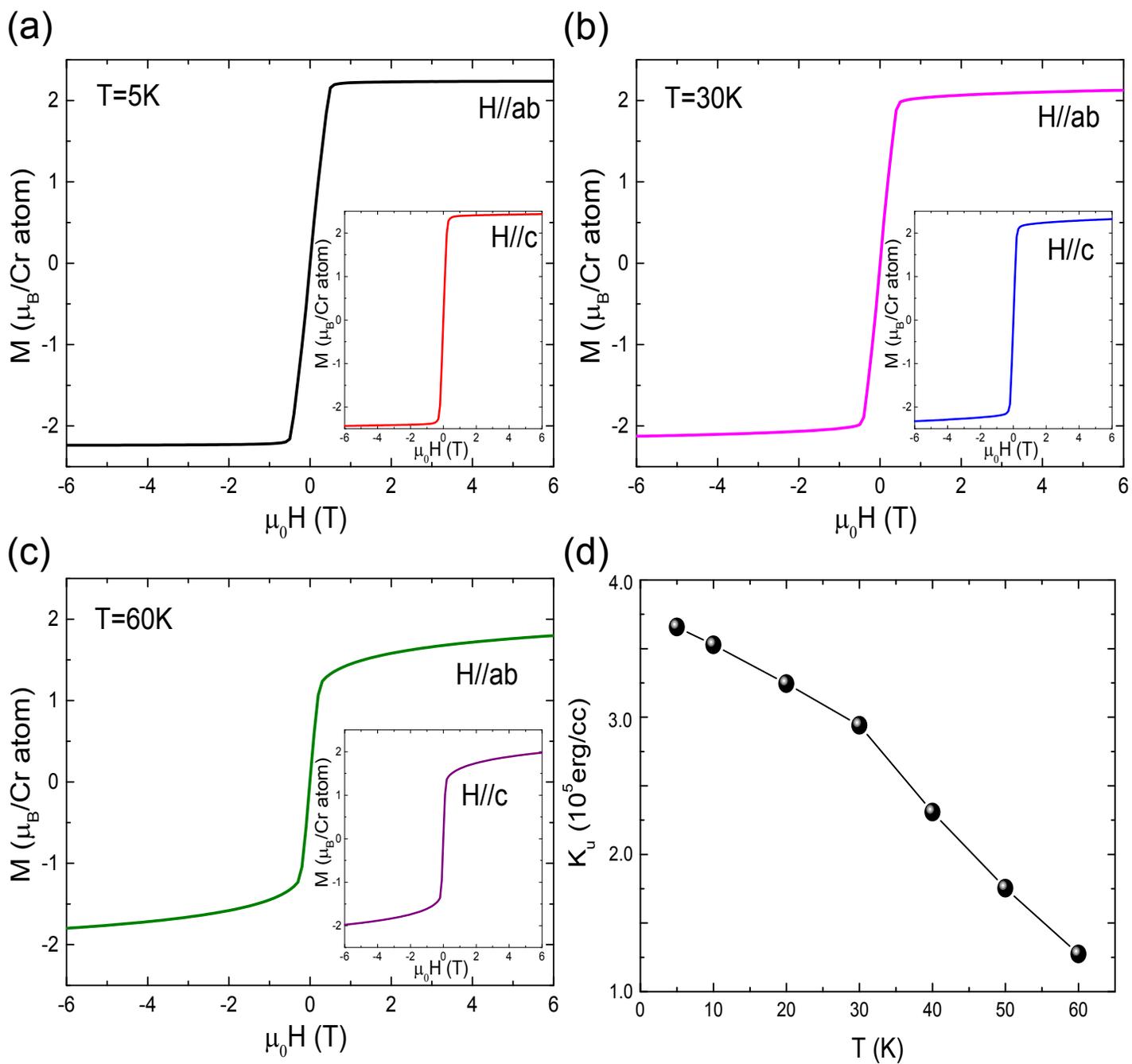